\documentclass[usenatbib,usegraphicx,twocolumn]{mn2e}

\newcommand{\vt}{{\vec \theta}}
\newcommand{\vr}{{\vec r}}
\newcommand{\vU}{{\vec{ U}}}

\newcommand{\Vo}{{\cal V}_{0}(\vec{U})}
\newcommand{\XI}{{\chi}(\vt)}
\newcommand{\tXI}{{\tilde{\chi}}(\vec{U})}
\newcommand{\Vi}{{\cal V}_{1}(\vec{U})}
\newcommand{\Po}{{\cal P}_{0}(\vec{U})}

\newcommand{\PIT}{{\cal P}_{1}^{T}(\vec{U})}

\newcommand{\PXTM}{{\cal P}_{\chi_{mod}}^{T}(\vec{U})}
\newcommand{\PXT}{{\cal P}_{\chi}^{T}(\vec{U})}
\newcommand{\PX}{{\cal P}_{\chi}(\vec{U})}

\newcommand{\PvT}{{\cal P}_{v}^{T}(\vec{U})}
\newcommand{\PvTM}{{\cal P}_{v_{mod}}^{T}(\vec{U})}

\def\L{{\cal L}}

\def\P{{\cal P}}

\def\HI{{\rm H~{\sc i}~}}

\usepackage{epsfig}
\usepackage{graphicx}
\usepackage{amsmath}

\begin{document}
\title{Visibility moments and power spectrum of turbulence velocity}
\author[Prasun Dutta]
{Prasun Dutta$^{1}$\thanks{Email:prasun@iiserb.ac.in},  
\\$^{1}$ IISER Bhopal, ITI Gas Rahat Building, Bhopal-462023, India.}

\date{}
\maketitle
\begin{abstract}
Here we introduce moments of visibility function and discuss how those can be used to estimate the power spectrum of the turbulent velocity of external spiral galaxies. We perform numerical simulation to confirm the credibility of this method and found that for galaxies with lower inclination angles it works fine. The estimator outlined here is unbiased and has the potential to recover the turbulent velocity spectrum completely from radio interferometric observations.
\end{abstract}

\begin{keywords}
physical data and process: turbulence-galaxy:disc-galaxies:ISM
\end{keywords}

\section{Introduction}
It is found that the power spectra of the \HI column density fluctuations in our Galaxy \citep{1983A&A...122..282C, 1993MNRAS.262..327G} follow power law  at length scales ranging a few parsec to few hundred parsecs.  Different probes to the statistics of the turbulent velocity fluctuations of our Galaxy; statistics of centroid of velocities \citep{2009RMxAC..36...45E}, velocity coordinate spectrum or VCS \citep{2009RMxAC..36...54P}, velocity channel analysis or VCA \citep{2000ApJ...537..720L}, spectral correlation function \citep{2001ApJ...555L..33P} show  that at  similar scales, the power spectrum of the velocity fluctuations also follow power laws. Theoretically, these structures are understood in terms of compressible fluid turbulence \citep{2004ARA&A..42..211E} driven by supernovae shocks, self gravity etc.  Recently, \HI column density power spectrum from external dwarf \citep{2006MNRAS.372L..33B, 2008MNRAS.384L..34D} and spiral galaxies \citep{2008MNRAS.384L..34D, 2009MNRAS.397L..60D, 2013NewA...19...89D} are also found to follow power law at  length scales ranging from a few parsec to $\sim 10$ kpc. 
Generating mechanism of these large scale structures are yet to be understood \citep{2014MNRAS.441..525W}. Estimating the velocity fluctuation power spectrum  would give more clue here. For this purpose, VCA is the most efficient technique and has been used by several authors successfully for our galaxy and nearby dwarf galaxies \citep{2001ApJ...551L..53S,2015ApJ...810...33C}.  However, for the external spiral galaxies, this technique has limited use \citep{2015MNRAS.452..803D}.

In this letter we propose to use moments of the observed visibilities to perform a model dependent estimate of the statistics of the line of sight component of turbulent velocities from external galaxies. We outline an estimator for the power spectrum of the turbulent velocity fluctuations and  access it's efficacy for a noise free simulated data.  
Section (2) defines visibility moments and  the estimator. Simulation parameters are given  in section (3) and in section (4) we explain the procedure to use the estimator on the simulated data. We discuss the results  and conclude in the last section (5).

\section{Power spectrum of visibility moments}
Specific intensity of radiation with rest frequency $\nu_{0}$ from \HI gas at $\vec{r}= (x, y, z) = (\vec{R}, z)$ at a distance $D$ and  having temperature $T$ can be written for small optical depth limit as  \citep{2011piim.book.....D}
\begin{equation}
I (\vec{\theta}, v)\ =\ I_{0}\,  \int d\,  z\, n_{HI} (\vec{r})\, \phi (v).
\end{equation}
Here $v = c ( \nu_{0} - \nu)/nu_{0}$,  $\nu$ is the frequency of observation,  $I_{0} = \frac{3 h \nu_{0} A_{21}}{16 \pi}$ and $z$ is the line of sight direction. $\vec{R}$ is in the plane of the sky, angular separation of a direction in the field of view with respect to the field centre is $\vec{\theta} = \frac{\vec{R}}{D}$. The line shape function $\phi(v)$ can be written as
\begin{equation}
\phi(v) = \phi_{0}\, \exp \left [ - \frac{ \left ( v - v_{z} (\vec{r}) \right ) ^{2} } {2 \sigma^{2}(\vec{r}) } \right ],
\end{equation}
were $\sigma(\vec{r}) = \sqrt{\frac{ k_{b} T}{m_{HI}}}$ \footnote{$k_{b}$ : Boltzmann constant, $m_{HI}$ : mass of hydrogen atom} is the thermal velocity dispersion and $v_{z}(\vec{r})$ is the line of sight velocity component of the \HI gas.



We use radio interferometers to observe \HI emissions from the  spiral galaxies.
Every antenna pair of an interferometric array measures the quantity termed as  visibility $V(\vU, v)$ at a baseline $\vU$ given by the projected antenna separation at the plane of the sky in units of observing wavelength.  Visibilities are direct Fourier transform of the specific intensity distribution, i.e,
\begin{equation}
V(\vU, v) \ =\ \int d \vt\  e^{2 \pi i \vt . \vU} I (\vec{\theta}, v).
\end{equation}
Depending on the antenna positions of the particular telescope array, integration time of observation and declination of the source, visibilities can only be measured at a finite set of  points in the $\vU$ plane. While reconstructing the image by performing an inverse Fourier transform of the observed visibilities, above sampling results in a complicated telescope beam pattern, usually known as dirty beam \citep{1999ASPC..180.....T}. Hence, image obtained by an inverse Fourier transform of the visibilities is a convolution of the dirty beam with the sky specific intensity. Various  deconvolution techniques are used  to estimate the specific intensity from the image.  However, all  these algorithms effectively interpolate the visibilities in the unsampled part of the  $\vU$ plane. This gives rise to  an unknown but  correlated noise   in the reconstructed image \citep{2011arXiv1102.4419D}. Hence, any estimate of power spectrum of  column density or  line of sight velocity,  evaluated using the reconstructed specific intensity is expected to  have an unknown bias.

We define the zeroth moment of the visibility function as follows
 \begin{equation}
\Vo = \int \ dv\  V (\vU, v). 
\end{equation}
The  velocity  integral above has to be performed over the entire spectral range of \HI emission. Clearly the moment $\Vo $ is proportional to the the Fourier transform of columns density $N_{HI}(\vt) = \int dz\, n_{HI}(\vec{R}, z)$. In fact, the quantity $\Po = \langle \mid \Vo \Vo^{*} \mid \rangle $ has been used in literature \citep{1993MNRAS.262..327G, 2013MNRAS.436L..49D} to estimate the power spectrum of the column density fluctuations of our Galaxy and external galaxies. We define the first moment of visibility function as
\begin{eqnarray}
 \Vi &=& \int \ dv\  v\ V (\vU, v) \\ \nonumber
 &=& \int d \vt\ e^{2 \pi i \vt . \vU}\  I_{0}\,  \int d\,  z\, n_{HI} (\vec{r})\, v_{z}(\vr).
\end{eqnarray}
Here $v_{z} (\vr)$, the  line of sight velocity, has two components:  $v^{\Omega}_{z}(\vt)$,  systematic rotation of the galaxy,  and  $v^{T}_{z}(\vt)$,  random motion because of turbulence. Hence, $v_{z}(\vr) = v_{z}^{\Omega}(\vr) + v_{z}^{T}(\vr)$. In case of the spiral galaxies with relatively small inclination angle, we may assume that the line of site component of the rotational velocity is independent of the distance along the line of sight, then
\begin{equation}
\Vi \ =\ \Vo \otimes  \tilde{v_{z}^{\Omega}} (\vU) \ +\ \tXI,
\end{equation}
where $\tilde{v_{z}^{\Omega}} (\vU)$ is the Fourier transform of $v_{z}^{\Omega}(\vr)$ and $\tXI$ is the Fourier transform of 
\begin{equation}
\XI = I_{0}\ \int dz\  n_{HI}(\vr) v_{z}^{T}(\vr).
\end{equation}

\subsection{An estimator of the turbulent velocity fluctuations}
It is safe to assume that the systematic rotational velocity of the galaxy and the random turbulent velocity are not correlated. We define 
\begin{equation}
\PX =\langle \mid \tXI \mid^{2} \rangle =\langle  \mid \Vi \mid^{2} -  \Po \otimes \mid \tilde{v_{z}^{\Omega}} (\vU)  \mid ^{2} \ \rangle.
\end{equation}
The angular brackets $< ... >$ denote ensemble averaging, which has to be taken over many statistical  realisations of the observed sky. As we are provided with only one realisation in reality, this can not be performed. Turbulence generated fluctuations in density and velocity, particularly for magnetohydrodynamic (MHD) turbulence, can have preffered orientations \citep{2005ApJ...631..320E}. However, it is not clear, if at the length scales of $1$ kpc to $10$ kpc or higher one would expect to see anisotropic fluctuations in neutral hydrogen resulting because of MHD turbulence. Here, we assume that we are probing homogenious and isotropic fluctuations only. The method presented here can be generalized for the anitropic case, however, observations with higher signal to noise than presently available may be required.  With this assumptions, the ensemble average in eqn.~(8) can be  replaced by averages over annular bins in  $\vU$. Note that the quantities $ \mid \Vi \mid^{2}$  and $\Po$ can be estimated directly from the observed visibilities. The line of sight component of the rotational velocity $v^{\Omega}_{z}(\vt)$, inclination or position angle etc depends only on the galacto-centric radius.  These quantities,  and hence also $\mid \tilde{v_{z}^{\Omega}} (\vU)  \mid $, can be estimated from the reconstructed image without any bias. It is clear from eqn.~(7) that $\PX$ contain in it informations about  \HI density as well as \HI velocity fluctuation power spectra, whereas the \HI density fluctuation power spectra can be estimated independently using $\Po$.

Next we assume that the turbulent velocity and the density fluctuations are not correlated and refer to the following two cases.

\begin{itemize}
\item If we are interested in estimating the power spectra at length scales comparable to the scale of the galaxy's disk, i.e, scales much larger than the scale height of the galaxy in consideration, we may approximate the galaxy as a thin disk. In this case, we can write 
\begin{equation}
\PX \ =\Po \otimes \PvT,
\end{equation}
where $\PvT = \langle \mid  v_{z}^{T} (\vU) \mid^{2} \rangle$ is the power spectrum of the turbulence velocity. It is to be noted that the quantities $\Po$ and $\PvT$ are two dimensional power spectra of the column density and the velocity fluctuations respectively. We have already discussed that $\Po$ can be measured from the zeroth moment of the observed visibilities. Since we expect all three functions in the above expression to be smooth and continuous, we can estimate  $\PvT$  either by deconvolving $\Po$ from $\PX$ or by regression analysis using a parametric model for $\PvT$. Since we have an independent and exact measure of $\Po$ from the zeroth moment of visibilities, we can have  a complete estimate of $\PvT$ in this case.

\item On the other hand, if we are to estimate the power spectra at scales comparable to the scale height of the galaxy's disk, we may write,
\begin{equation}
\PX \ = P^{3D}_{n_{HI}} (\vU, 0) \otimes P^{3D}_{v^{T}_{z}}(\vU, 0),
\end{equation}
where $P^{3D}_{n_{HI}} (\vU, 0)$ and $P^{3D}_{v^{T}_{z}}(\vU, 0)$ are the three dimensional power spectra of $n_{HI}(\vr)$ and $v^{T}_{z}(\vr)$ evaluated at the zero line of sight wavenumber. It can be shown that $P^{3D}_{n_{HI}} (\vU, 0)$ is  proportional to  $\Po$ for a thick disk. Hence, just like the  case for the thin disk, we may use $\Po$ to find the turbulent velocity fluctuation power spectrum from above relation either by deconvolution or by regression analysis. However, our ability to estimate the amplitude of $P^{3D}_{v^{T}_{z}}$ would depend on the details of the morphology of the galaxy in the line of sight direction. 
\end{itemize}

\section{Simulating \HI observations from a spiral galaxy}

\begin{figure}
\begin{center}
\epsfig{file=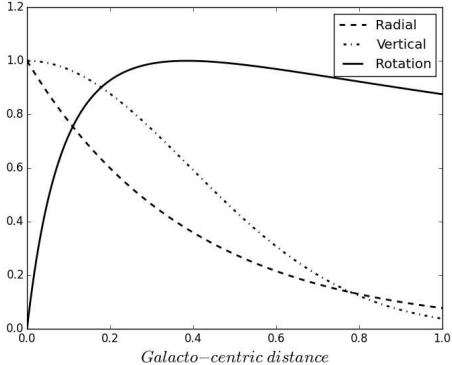,width=2.8in}
\end{center}
\caption{Radial profile (dash line), vertical profile (dot-dash line) of the \HI number density  adopted for simulation are plotted against distance from the centre of the galaxy in radial and vertical directions. The solid black line represents the rotation curve used to generate $v^{\Omega}_{z}(\vt)$. Note distances are normalised to the size of the simulation box.}
\label{fig:profile}
\end{figure}
A major assumption we took in deriving the expression for $\PX$  in the previous section is that $v^{\Omega}_{z}$ is independent of the line of sight direction. In case of galaxies with lower inclination angle, this assumption is justified. However, for  galaxies with higher inclination angles, line of sight pass through different galacto-centric radius and this assumption may not be justified.
We simulate   \HI observation from an external  spiral  galaxy and use it to  study the effect of galaxy's inclination on the turbulent velocity power spectrum estimator  described in the previous section. Simulation of the \HI signal  is based on eqn.~(1) and (2) and discussed in \citet{2015MNRAS.452..803D} in detail, we note the salient features here.  We consider an exponential radial profile and a gaussian vertical profile to model the \HI distribution in the galaxy, i.e,
\begin{equation}
\begin{aligned}
n_{HI} (\vec{r_{g}})  \ =\ n_{0}\  & \exp\left (- \mid \vec{R}_{g} \mid / R_{0}  \right )\  \exp \left ( -\frac{1}{2} \left( z/z_{0} \right )^2 \right ) \\
&\left [ 1 + f_{HI}\ \delta n_{HI} (\vec{r_{g}}) \right ],
\end{aligned}
\end{equation}
where $\vec{r_{g}} = (\vec{R{g}}, z_{g})$ is a vector centred at the galactic centre with $\vec{R_{g}}$ at the plane of the galaxy. We assume a constant inclination angle $i$ between the line of sight direction and $z_{g}$ for simplicity. This gives a rotation mapping between the coordinates $\vec{r}$ and $\vec{r}_{g}$. Turbulent \HI number density fluctuation $\delta_{HI}(\vec{r}_{g})$ is expected to have a power law spectrum and we quantify it with a slope $\alpha$. We assume that the tangential rotation velocity $v_{t}(R_{g})$ follow Brandt profile \citep{1960ApJ...131..293B} of the form  
\begin{equation}
v_{t}(R_{g}) \ =\ \frac{ \frac{R_{g} } {R_{v}}  v_{0} }{ \left [  \frac{1}{3} + \frac{2}{3} \frac{R_{g}}{R_{v}} \right ]^{3/2} },
\end{equation}
and use this to generate the line of sight component of the systematic velocity of the \HI gas across the galactic disk 
\begin{equation}
v_{z}(\vec{r}) \ =\ \frac{v_{t}(R_{g}) sin(i) x }{R_{g}} +  f_{v}v^{T}_{z}(\vec{r}).
\end{equation}
We shall discuss the other parameters in above expressions shortly.

We consider $512^{3}$ cubic grids in our simulation volume with each cube representing an individual \HI cloud. To accommodate the simulated galaxy's disk inside the simulation volume and adopt a generic rotation curve, we considered both $R_{0} = R_{v} = 100$ grid units. Assuming \HI extent of a typical  spiral galaxy is $\sim 50$ kpc these choice gives us  a spatial resolution of $\sim 200$ pc.  Simulating a thin disk requires relatively higher computational resources. On the other hand inclination angle affects both the thick and thin disks in similar way, we opted for simulating a thick disk here and  adopt a value of $z_{0} = 100$ grid units.  We show these radial and vertical profiles normalised to their central value in Figure~(1) along with the tangential velocity profile normalised to $v_{0}$. 

\begin{figure}
\begin{center}
\epsfig{file=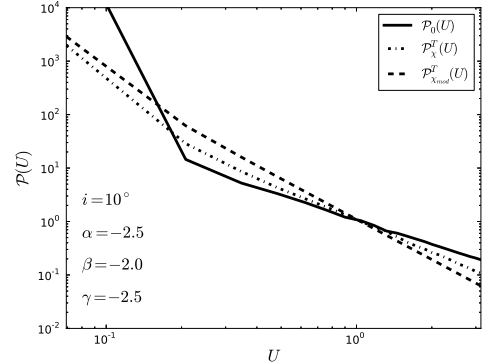,width=2.8in}
\end{center}
\caption{$\PIT$ estimated from the simulation is shown with dot-dashed line. This uses the parameters given in the bottom left corner of the image ($\beta = -2.0$). A model spectrum $\PXTM$ with $\gamma = -2.5$ is also shown (dashed line) and it does not match with $\PXT$. We also show the corresponding column density power spectrum  (solid black line). All three spectra are normalised at $U=1$.}
\label{fig:PS}
\end{figure}

\begin{figure*}
\begin{center}
\epsfig{file=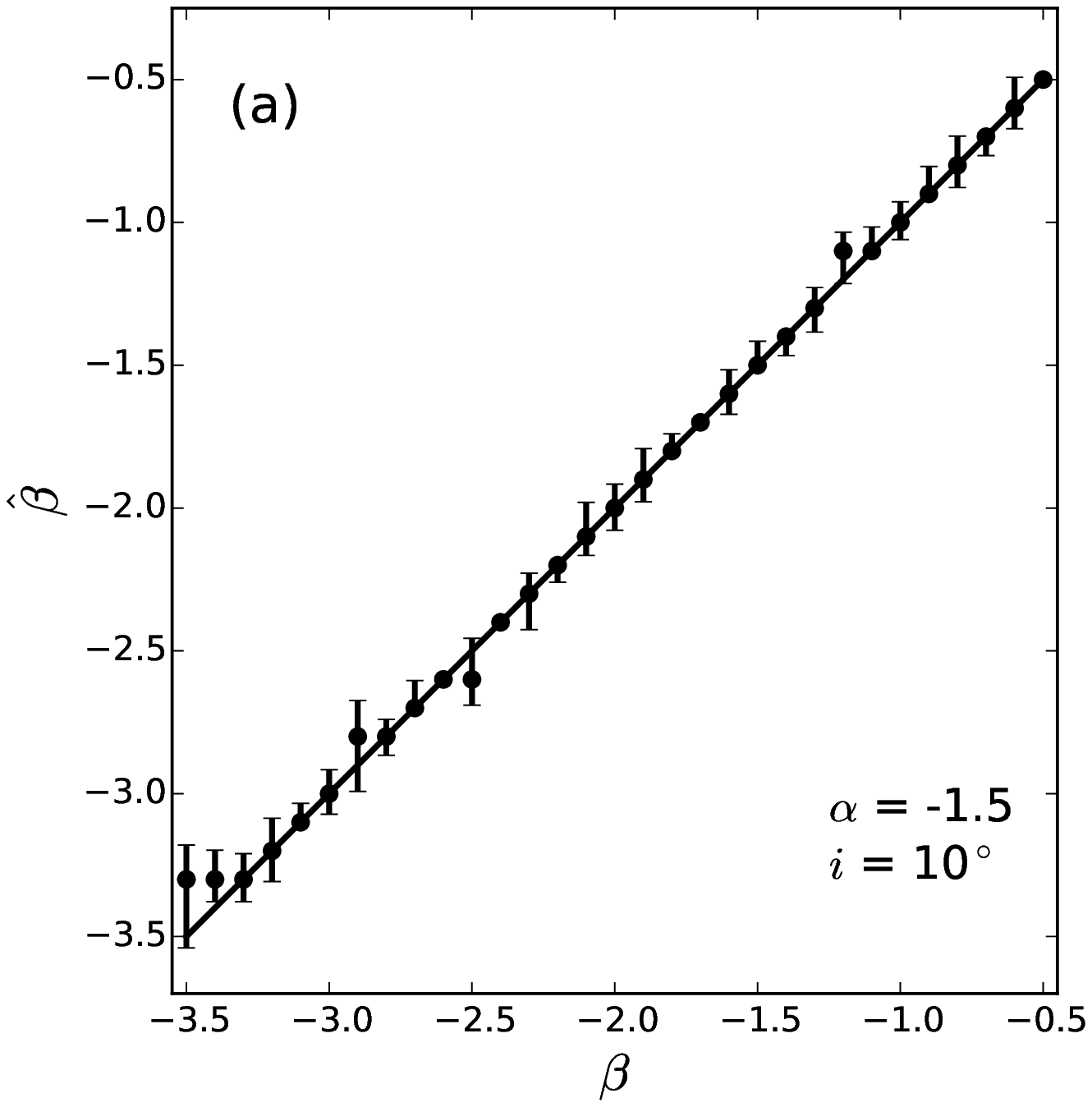,width=2.4in}
\epsfig{file=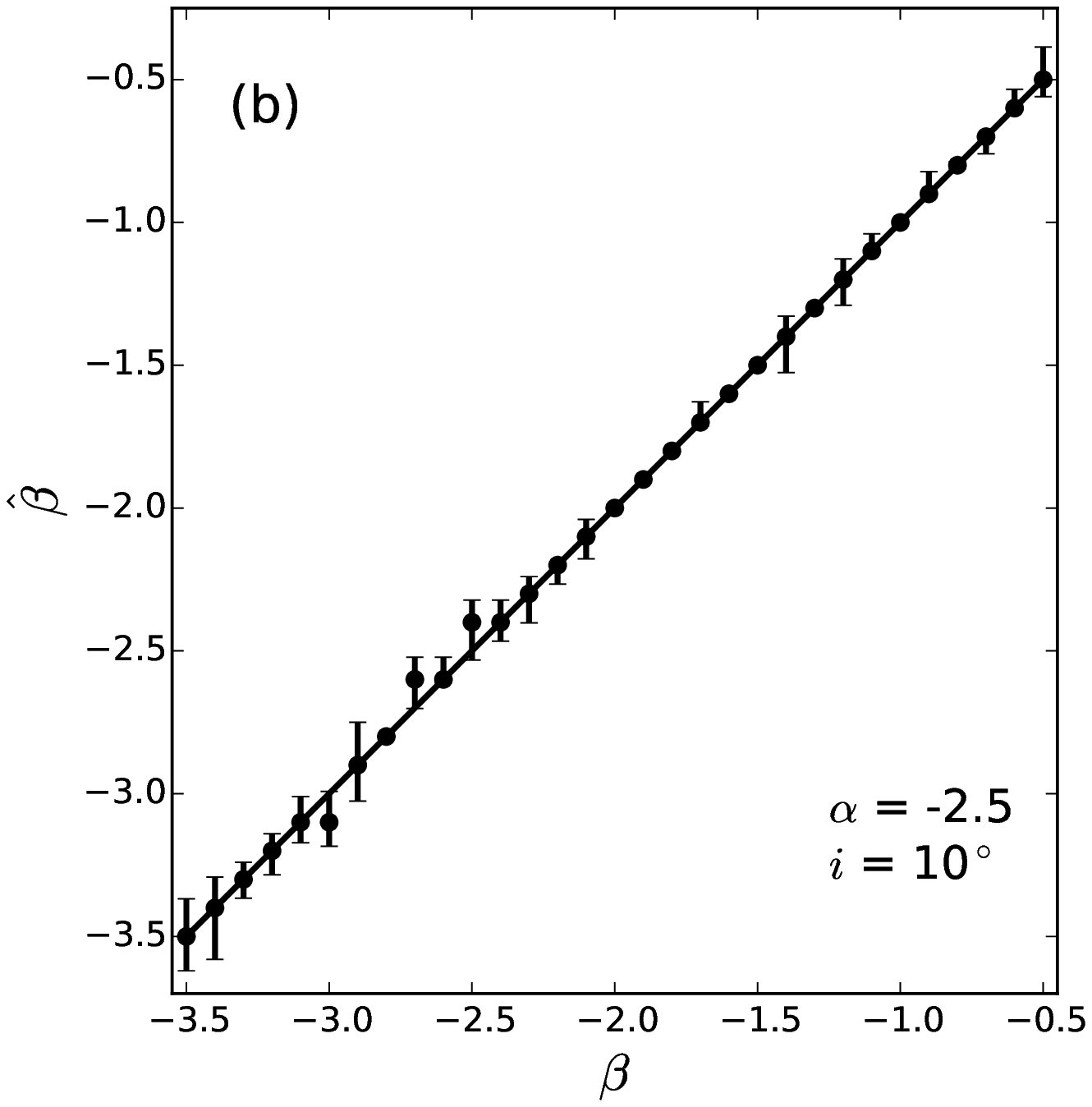,width=2.4in}\\
\epsfig{file=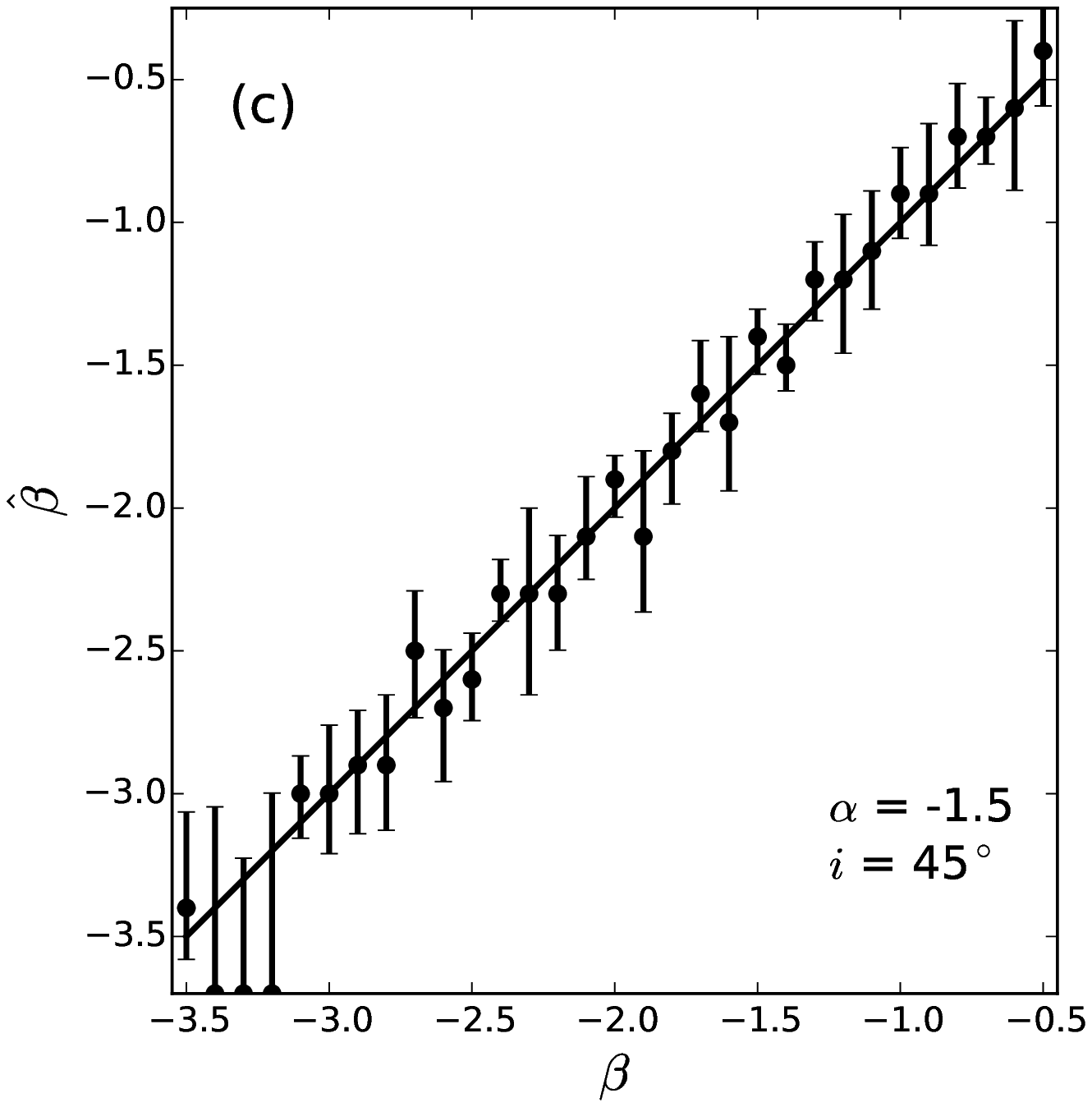,width=2.4in}
\epsfig{file=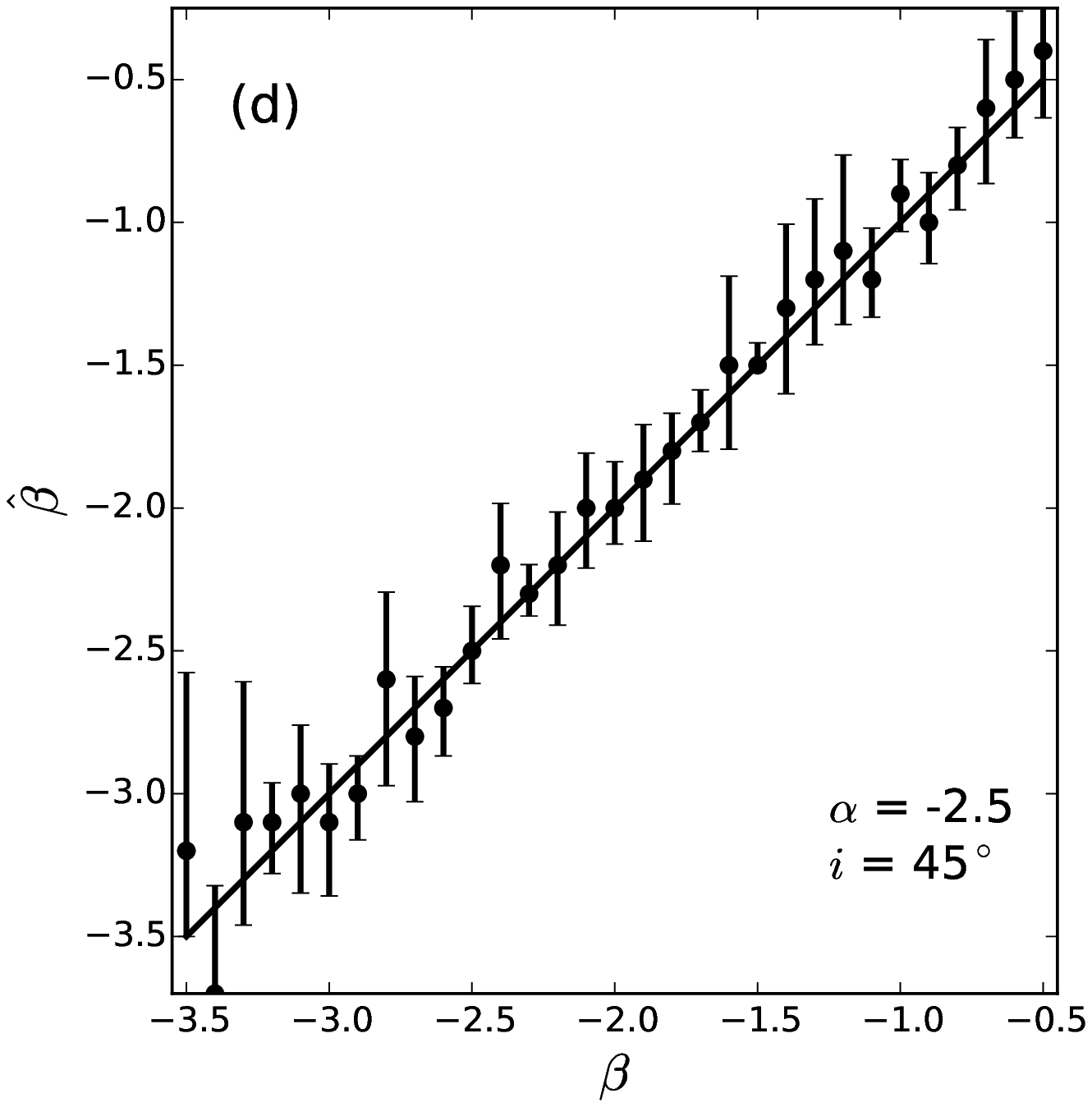,width=2.4in}
\end{center}
\caption{Best fit values of $\beta$, i.e, $\hat{\beta}$  found using the estimator to the simulated data is plotted against the input values of $\beta$ with $3\sigma$ errors for four different sets of simulations. The solid line represents the ideal case of $\hat{\beta} = \beta$. Clearly at high inclination angles the estimated values of $\beta$ has relatively higher uncertainty.}
\label{fig:NHI}
\end{figure*}

 \citet{2013MNRAS.436L..49D} have reported that the amplitude of the \HI column density fluctuation of the spiral galaxies in the  THINGS sample\footnote{THINGS: The \HI Nearby Galaxy Survey \citep{2008AJ....136.2563W}} are approximately $1/10^{th}$ of the mean column density. Following this, we choose $f_{HI} = 0.1$. Line of sight velocity dispersion of \HI for the galaxies in the THINGS sample varies in the range  $\sim 5 $ to $20$ km sec$^{-1}$ \citep{2009AJ....137.4424T}, while the flattening value of the rotation curve are found to  be in the range $\sim 100$ to $200$ km sec$^{-1}$ \citep{2008AJ....136.2648D}. Since this velocity dispersion is mostly because of turbulence, we adopt $f_{v} = v_{0}/10$ in the simulation,  the actual value of $v_{0}$ is unimportant here. \citet{2011ApJ...737L..33K} found that in our Galaxy, line of sights with column density $> 2 \times 10^{20}$ atoms cm$^{2}$ have temperature of $240$ K or lower, whereas the mean column density of the spiral galaxies in the THINGS sample is $> 3 \times 10^{20}$ atoms cm$^{2}$ \citep{2008AJ....136.2563W,2013MNRAS.436L..49D}. This suggest existence of large amount of cold gas in the spiral galaxies. In view of this, thermal velocity dispersion $\sigma$ is considered to be $v_{0}/100$. For a $ v_{0} = 200 $ km sec$^{-1}$, this corresponds to a gas  temperature $500$ K, i.e from cold neutral medium.
 
 \HI column density power spectra of 18 spiral galaxies in the THINGS sample  are found to follow power laws \citep{2013NewA...19...89D}  for length scales more than the scale height of the disk and the slope of the power law for most of the galaxies were found to be in the range $-1.5$ to $-1.8$.   To see the effect of $\alpha$ on the efficacy of our estimator we perform the simulation for two different  $\alpha = (-2.5, -1.5)$.  We consider a nearly face on disk with $i = 10^{\circ}$ and a more edge on disk with $i = 45 ^{\circ}$. 
With these choices,  we run four sets of simulations as {\bf  (a)} $( i = 10^{\circ}, \alpha = -1.5)$, {\bf (b)} $( i = 10^{\circ}, \alpha = -2.5)$, {\bf (c)} $( i = 45^{\circ}, \alpha = -1.5)$ and {\bf (d)} $( i = 45^{\circ}, \alpha = -2.5)$. Turbulent velocity component $v^{T}(\vec{r})$ is also taken to follow power law power spectrum \citep{2009ApJ...692..364F},  with a slope $\beta$.  \citet{2000ApJ...541..701G}  pointed out that it is likely that the ISM density fluctuations passively follow the velocity fluctuations and hence they would have similar power spectra. Observations  show that the column density fluctuation power spectra has a slope of $\sim -2.5$ \citep{1993MNRAS.262..327G} in our Galaxy and  $\sim -1.5$  \citep{2013MNRAS.436L..49D} in case of the external galaxies. We simulate the \HI emission for $\beta$ varying in the range $-0.5$ to $-3.5$ in steps of $0.1$ and generate \HI specific intensity data cubes following eqn.~(1) with a velocity resolution of $\sigma/2$. This amounts in calculating the line shape function  for at least six points within it's support, necessary and sufficient to avoid the shot noise bias discussed in \citet{2009ApJ...693.1074C}. We convert these specific intensity maps to corresponding visibilities and use them as the observed visibilities $V(\vU, v)$ for further analysis. In this letter  we do not consider any measurement noise. Further, visibilities are  estimated at grids in $\vec{U}$ space and hence the baseline coverage is assumed to be complete.

\section{Estimating turbulent velocity fluctuation from the simulated data}


We proceed to estimate the power spectrum of turbulent velocity fluctuations through the following steps.
\begin{itemize}
\item
We estimate the visibility moments $\Vo$ and $\Vi$ from the visibilities $V(\vU, v)$.
\item
We reconstruct the image data cube from visibilities and estimate the rotation curve using a standard tilted ring fit to the data. Since, in the simulation we have kept the inclination angle constant, we also keep it to be constant with radius while performing the tilted ring fit. For a real data, we would need to consider the uncertainties in the estimate of the rotation curve (and also inclination and position angles). At the present case these uncertainties are not considered. Using the rotation curve estimate we construct $\tilde{v}^{\Omega}_{z}(\vU)$.
\item
We bin the data in annular bins and estimate the azimuthally averaged $\P_{0}(U)$ from the visibility moment $\Vo$ and $\PX$ from $\Vi$ and  $\tilde{v}^{\Omega}_{z}(\vU)$ in the same bins. 
\item
Assuming the power spectrum of the turbulent velocity fluctuations to follow power law, we consider $\PvTM \sim U^{\gamma}$ to model its power spectrum. We convolve $\PvTM$ with $\Po$ to make the azimuthally averaged model $\PXTM$. It is important to realise here that as discussed in Section~(3) our model of $\PXTM$ derived this way does not have the proper amplitude. We shall restrict ourself in estimating the slope of the turbulent velocity power spectra here. The solid line in Figure~(2) shows the column density power spectra for the case (b). The dotted line corresponds to $\PXT$ for $\beta = -2.0$ while the dot-dash line shows the model $\PXTM$ with $\gamma=-2.5$. As expected, the model and the estimated value are in mismatch here.

\item 
We vary $\gamma$ and compare $\PXTM$ with   $\PXT$ by a least square estimator with likelihood function of the form $\exp(-\L)$, where $\L$ is defined as
\begin{equation}
\L(\gamma) = \frac{L (\gamma)}{L_{min}},\ \ L(\gamma) =\ \sum_{i = 1}^{N} \left [  \mathcal{P}_{\chi}^{T} \left (\frac{U}{U_{0}} \right )[i]  - \mathcal{P}_{\chi_{ mod}}^{T} \left ( \frac{U[i]}{U_{0}} \right ) \right ]^{2}.
\end{equation}
Here the index `i' refer to a particular annular  bin and $N$ is the total number of bins.  Note that both the power spectra are normalised at a given baseline $U_{0}$ and hence this estimator is not sensitive to the amplitude of the velocity fluctuation spectra.  We find the best fit value of $\gamma$, i.e, $\hat{\beta}$ and also error in estimating this value.

\end{itemize}
 
Figure~(3) summarises the  efficacy of the model dependent maximum likelihood estimation of the slope of the turbulent velocity fluctuation power spectrum from the four sets of simulations. Let us concentrate on the panel (a) first. Here $\hat{\beta}$ is plotted against the different values of input $\beta$ for $\alpha = -1.5$ and $i = 10^{\circ}$ with $3\sigma$ error bars. The solid line corresponds to an exact match between the input and estimated values of $\beta$. 

Clearly, for a lower inclination angle, i.e $i = 10^{\circ}$, the estimator does a better job in accessing the real value of $\beta$. For the higher inclination angle,  i.e $i = 45^{\circ}$, for all the values of $\beta$ the best estimate has systematically higher uncertaininity. This is expected as for high inclination angle our assumption of independency of $v_{z}$ on line of sight coordinate is poorly valid. We also see for $\beta$ values $< -3.0$ the measurement uncertainty is considerably high.

\section{Discussions and Conclusions}
In this letter we define visibility moments and  an estimator of turbulent velocity fluctuations based on the visibility moments.Using  numerical simulations we find that  for lower inclination angles and  turbulent velocity fluctuation power law slope  $>-3.0$, the visibility moment estimator can  be used.  Though we have used a model dependent approach here, in principle it is possible to deconvolve the column density power spectra from $\PXT$ and measure the  velocity power spectrum of the line of sight velocity fluctuations directly. Moreover, applying  this method to  the   thin disk case, we can recover the turbulent velocity power spectrum completely.  This would be particularly useful to access the energy involved in the large scale fluctuations seen in external galaxies.

We have considered several simplifying assumptions. Firstly, we  ignored the uncertainty in reconstructing the  rotation velocity of the galaxy from the data, observational noise and incomplete baseline coverage. Each of these are expected to provide additional uncertainty for a real data and need to be looked upon carefully. Particularly, recovering the turbulent velocity power spectra depends highly on the accuracy of the rotational velocity estimate. For galaxies with warp disks, estimating rotational velocity is tricky and this method may not be useful there. Random fluctuations that may arise because of different systematic effects in the galactic rotation, however, are not expected to have scale invariant nature.  An unwanted mixer of these fluctuations with the fluctuations generated because of the underlying turbulence would destroy the power law nature of the spectrum.  Hence, estimate of the velocity spectra itself can be used as a self diagnostic test. 

The approach described here to estimate the line of sight turbulent velocity fluctautions from the external galaxies, apparently follow the same philosophy from the method of statistics of centroids of velocities \citep{2005ApJ...631..320E}, where  the unnormalised first moment of the specific intensity function is used to estimate  turbulent velocity statistics. Main difference in our approach is that we use the visibility function directly. Hence, it  does not have any unwanted noise bias like that arise because of  image reconstruction from incompletely sampled visibilities. 

Scaling relations of compressible and highly magnetized turbulence are studied in \citet{2007ApJ...666L..69K}. \citet{2005ApJ...631..320E} has also discussed incompatibility of the velocity centroid method for high Mach number turbulence. Though the $10$ kpc scales coherent structures seen at external galaxies are most likely be the result of ISM turbulence, it is neither clear if the neutral medium (\HI) at these scales would have an MHD coupling or they are highly supersonic. In either case, the applicability of the method described here needs to be investigated. Another basic assumption in developing this method has been that the \HI is optically thin. The interpretation of the intensity spectra would change in case the optical depth of the \HI is sufficiently higher \citep{2004ApJ...616..943L}. 

Nevertheless, here we discussed an  effective way of measuring the statistics of the turbulent velocity fluctuations from external galaxies at large scales.
We plan to use this estimator to the spiral galaxies with  lower inclination angle from the THINGS sample in near future.

\section*{Acknowledgement}
PD would like to acknowledge useful discussion with Nirupam Roy. This work is supported by the DST INSPIRE Faculty Fellowship award  [IFA-13 PH 54] and performed at Indian Institute of Science Education and Research, Bhopal. 
PD is thankful to Swagat Saurav Mishra for reading the earlier version of the draft and providing valuable comments.

\bibliographystyle{mn2e}
\bibliography{references}

\begin{thebibliography}{}

\bibitem[\protect\citeauthoryear{{Begum}, {Chengalur} \& {Bhardwaj}}{{Begum}
  et~al.}{2006}]{2006MNRAS.372L..33B}
{Begum} A.,  {Chengalur} J.~N.,    {Bhardwaj} S.,  2006, \mnras, 372, L33

\bibitem[\protect\citeauthoryear{{Brandt}}{{Brandt}}{1960}]{1960ApJ...131..293B}
{Brandt} J.~C.,  1960, \apj, 131, 293

\bibitem[\protect\citeauthoryear{{Chepurnov}, {Burkhart}, {Lazarian} \&
  {Stanimirovic}}{{Chepurnov} et~al.}{2015}]{2015ApJ...810...33C}
{Chepurnov} A.,  {Burkhart} B.,  {Lazarian} A.,    {Stanimirovic} S.,  2015,
  \apj, 810, 33

\bibitem[\protect\citeauthoryear{{Chepurnov} \& {Lazarian}}{{Chepurnov} \&
  {Lazarian}}{2009}]{2009ApJ...693.1074C}
{Chepurnov} A.,  {Lazarian} A.,  2009, \apj, 693, 1074

\bibitem[\protect\citeauthoryear{{Crovisier} \& {Dickey}}{{Crovisier} \&
  {Dickey}}{1983}]{1983A&A...122..282C}
{Crovisier} J.,  {Dickey} J.~M.,  1983, \aap, 122, 282

\bibitem[\protect\citeauthoryear{{de Blok}, {Walter}, {Brinks}, {Trachternach},
  {Oh} \& {Kennicutt} Jr.}{{de Blok} et~al.}{2008}]{2008AJ....136.2648D}
{de Blok} W.~J.~G.,  {Walter} F.,  {Brinks} E.,  {Trachternach} C.,  {Oh}
  S.-H.,    {Kennicutt} Jr. R.~C.,  2008, \aj, 136, 2648

\bibitem[\protect\citeauthoryear{{Draine}}{{Draine}}{2011}]{2011piim.book.....D}
{Draine} B.~T.,  2011, {Physics of the Interstellar and Intergalactic Medium}

\bibitem[\protect\citeauthoryear{{Dutta}}{{Dutta}}{2011}]{2011arXiv1102.4419D}
{Dutta} P.,  2011, ArXiv e-prints : 1102.4419

\bibitem[\protect\citeauthoryear{{Dutta}}{{Dutta}}{2015}]{2015MNRAS.452..803D}
{Dutta} P.,  2015, \mnras, 452, 803

\bibitem[\protect\citeauthoryear{{Dutta}, {Begum}, {Bharadwaj} \&
  {Chengalur}}{{Dutta} et~al.}{2008}]{2008MNRAS.384L..34D}
{Dutta} P.,  {Begum} A.,  {Bharadwaj} S.,    {Chengalur} J.~N.,  2008, \mnras,
  384, L34

\bibitem[\protect\citeauthoryear{{Dutta}, {Begum}, {Bharadwaj} \&
  {Chengalur}}{{Dutta} et~al.}{2009}]{2009MNRAS.397L..60D}
{Dutta} P.,  {Begum} A.,  {Bharadwaj} S.,    {Chengalur} J.~N.,  2009, \mnras,
  397, L60

\bibitem[\protect\citeauthoryear{{Dutta}, {Begum}, {Bharadwaj} \&
  {Chengalur}}{{Dutta} et~al.}{2013}]{2013NewA...19...89D}
{Dutta} P.,  {Begum} A.,  {Bharadwaj} S.,    {Chengalur} J.~N.,  2013, \na, 19,
  89

\bibitem[\protect\citeauthoryear{{Dutta} \& {Bharadwaj}}{{Dutta} \&
  {Bharadwaj}}{2013}]{2013MNRAS.436L..49D}
{Dutta} P.,  {Bharadwaj} S.,  2013, \mnras, 436, L49

\bibitem[\protect\citeauthoryear{{Elmegreen} \& {Scalo}}{{Elmegreen} \&
  {Scalo}}{2004}]{2004ARA&A..42..211E}
{Elmegreen} B.~G.,  {Scalo} J.,  2004, \araa, 42, 211

\bibitem[\protect\citeauthoryear{{Esquivel} \& {Lazarian}}{{Esquivel} \&
  {Lazarian}}{2005}]{2005ApJ...631..320E}
{Esquivel} A.,  {Lazarian} A.,  2005, \apj, 631, 320

\bibitem[\protect\citeauthoryear{{Esquivel} \& {Lazarian}}{{Esquivel} \&
  {Lazarian}}{2009}]{2009RMxAC..36...45E}
{Esquivel} A.,  {Lazarian} A.,  2009, in Revista Mexicana de Astronomia y
  Astrofisica Conference Series Vol.~36 of Revista Mexicana de Astronomia y
  Astrofisica, vol. 27, {Statistics of Centroids of Velocity}.
pp 45--53

\bibitem[\protect\citeauthoryear{{Federrath}, {Klessen} \&
  {Schmidt}}{{Federrath} et~al.}{2009}]{2009ApJ...692..364F}
{Federrath} C.,  {Klessen} R.~S.,    {Schmidt} W.,  2009, \apj, 692, 364

\bibitem[\protect\citeauthoryear{{Goldman}}{{Goldman}}{2000}]{2000ApJ...541..701G}
{Goldman} I.,  2000, \apj, 541, 701

\bibitem[\protect\citeauthoryear{{Green}}{{Green}}{1993}]{1993MNRAS.262..327G}
{Green} D.~A.,  1993, \mnras, 262, 327

\bibitem[\protect\citeauthoryear{{Kanekar}, {Braun} \& {Roy}}{{Kanekar}
  et~al.}{2011}]{2011ApJ...737L..33K}
{Kanekar} N.,  {Braun} R.,    {Roy} N.,  2011, \apjl, 737, L33

\bibitem[\protect\citeauthoryear{{Kowal} \& {Lazarian}}{{Kowal} \&
  {Lazarian}}{2007}]{2007ApJ...666L..69K}
{Kowal} G.,  {Lazarian} A.,  2007, \apjl, 666, L69

\bibitem[\protect\citeauthoryear{{Lazarian} \& {Pogosyan}}{{Lazarian} \&
  {Pogosyan}}{2000}]{2000ApJ...537..720L}
{Lazarian} A.,  {Pogosyan} D.,  2000, \apj, 537, 720

\bibitem[\protect\citeauthoryear{{Lazarian} \& {Pogosyan}}{{Lazarian} \&
  {Pogosyan}}{2004}]{2004ApJ...616..943L}
{Lazarian} A.,  {Pogosyan} D.,  2004, \apj, 616, 943

\bibitem[\protect\citeauthoryear{{Padoan}, {Kim}, {Goodman} \&
  {Staveley-Smith}}{{Padoan} et~al.}{2001}]{2001ApJ...555L..33P}
{Padoan} P.,  {Kim} S.,  {Goodman} A.,    {Staveley-Smith} L.,  2001, \apjl,
  555, L33

\bibitem[\protect\citeauthoryear{{Pogosyan} \& {Lazarian}}{{Pogosyan} \&
  {Lazarian}}{2009}]{2009RMxAC..36...54P}
{Pogosyan} D.,  {Lazarian} A.,  2009, in Revista Mexicana de Astronomia y
  Astrofisica Conference Series Vol.~36 of Revista Mexicana de Astronomia y
  Astrofisica, vol. 27, {Line-of-sight statistical methods for turbulent
  medium: VCS for emission and absorption lines}.
pp 54--59

\bibitem[\protect\citeauthoryear{{Stanimirovi{\'c}} \&
  {Lazarian}}{{Stanimirovi{\'c}} \& {Lazarian}}{2001}]{2001ApJ...551L..53S}
{Stanimirovi{\'c}} S.,  {Lazarian} A.,  2001, \apjl, 551, L53

\bibitem[\protect\citeauthoryear{{Tamburro}, {Rix}, {Leroy}, {Mac Low},
  {Walter}, {Kennicutt}, {Brinks} \& {de Blok}}{{Tamburro}
  et~al.}{2009}]{2009AJ....137.4424T}
{Tamburro} D.,  {Rix} H.-W.,  {Leroy} A.~K.,  {Mac Low} M.-M.,  {Walter} F.,
  {Kennicutt} R.~C.,  {Brinks} E.,    {de Blok} W.~J.~G.,  2009, \aj, 137, 4424

\bibitem[\protect\citeauthoryear{{Taylor}, {Carilli} \& {Perley}}{{Taylor}
  et~al.}{1999}]{1999ASPC..180.....T}
{Taylor} G.~B.,  {Carilli} C.~L.,    {Perley} R.~A.,  eds, 1999, {Synthesis
  Imaging in Radio Astronomy II} Vol.~180 of Astronomical Society of the
  Pacific Conference Series

\bibitem[\protect\citeauthoryear{{Walker}, {Gibson}, {Pilkington}, {Brook},
  {Dutta}, {Stanimirovi{\'c}}, {Stinson} \& {Bailin}}{{Walker}
  et~al.}{2014}]{2014MNRAS.441..525W}
{Walker} A.~P.,  {Gibson} B.~K.,  {Pilkington} K.,  {Brook} C.~B.,  {Dutta} P.,
   {Stanimirovi{\'c}} S.,  {Stinson} G.~S.,    {Bailin} J.,  2014, \mnras, 441,
  525

\bibitem[\protect\citeauthoryear{{Walter}, {Brinks}, {de Blok}, {Bigiel},
  {Kennicutt} Jr., {Thornley} \& {Leroy}}{{Walter}
  et~al.}{2008}]{2008AJ....136.2563W}
{Walter} F.,  {Brinks} E.,  {de Blok} W.~J.~G.,  {Bigiel} F.,  {Kennicutt} Jr.
  R.~C.,  {Thornley} M.~D.,    {Leroy} A.,  2008, \aj, 136, 2563

\end{thebibliography}

\end{document}